\documentclass{svproc}

\usepackage{cite}
\usepackage{amsmath,amssymb,amsfonts}
\usepackage{algorithmic}
\usepackage{graphicx}
\usepackage{textcomp}
\usepackage{xcolor}
\usepackage{fullpage}
\usepackage{caption}
\usepackage{subcaption}
\usepackage{float}
\usepackage{comment}

\def\BibTeX{{\rm B\kern-.05em{\sc i\kern-.025em b}\kern-.08em
    T\kern-.1667em\lower.7ex\hbox{E}\kern-.125emX}}

\begin{document}

\title{Malware Analysis with Artificial Intelligence and a Particular Attention on Results Interpretability}


\author{Benjamin Marais\inst{1}\inst{2} \and Tony Quertier\inst{1} \and Christophe Chesneau\inst{2}}

\institute{Orange Labs, France\\
\email{benjamin.marais@orange.com, tony.quertier@orange.com}
\and
Department of Mathematics, LMNO, University of Caen Normandy, France\\
\email{christophe.chesneau@unicaen.fr}}

\authorrunning{Benjamin Marais et al.} 

\maketitle

\abstract{Malware detection and analysis are active research subjects in cybersecurity over the last years. Indeed, the development of obfuscation techniques, as packing, for example, requires special attention to detect recent variants of malware. The usual detection methods do not necessarily provide tools to interpret the results. Therefore, we propose a model based on the transformation of binary files into grayscale image, which achieves an accuracy rate of 88\%. Furthermore, the proposed model can determine if a sample is packed or encrypted with a precision of 85\%. It allows us to analyze results and act appropriately. Also, by applying attention mechanisms on detection models, we have the possibility to identify which part of the files looks suspicious. This kind of tool should be very useful for data analysts, it compensates for the lack of interpretability of the common detection models, and it can help to understand why some malicious files are undetected.}

\section{Introduction}
In recent years, the number of malware and attacks has increased exponentially. The illustration of this phenomenon is the number of online submissions to sandboxes, such as Virustotal or Any.run, among other things. In addition, these malware are increasingly difficult to detect due to very elaborate evasion techniques. For example, polymorphism is used to evade pattern-matching detection relied on by security solutions like antivirus software, while some characteristics of polymorphic malware change, its functional purpose remains the same. These developments make obsolete detection solutions like signature-based detection. Researchers and companies have therefore turned to artificial intelligence methods to manage both large volumes and complex malware. In this paper, we will look at the static analysis of malware for computational issues such as time and resources. Indeed, dynamic analysis gives very good results, but for companies that have thousands of suspicious files to process, it creates resource problems because a sandbox can require two to three minutes per file.

\subsection{State of art}

Malware detection and analysis represent very active fields of study. In recent years, several methods have been proposed in this regard.

The most popular detection method is signature-based detection \cite{sung2004static} \cite{sathyanarayan2008signature}. This method consists of stocking portions of code of benign and malicious files called signatures. It consists of comparing the signature of a suspicious file with the signature database. A weakness of this method is having the file first, determining its nature and recorded its signature.

Another common and effective method is called dynamic analysis. It attempts to run suspect files in secure environments (physical or virtual) named sandbox \cite{Vasilescu2014}. It allows analysts to study the behavior of the file without risk. This process is particularly effective in detecting new malware or known malware that has been modified with obfuscation techniques. This procedure, however, may be a waste of time and resources. Also, some malware is able to detect virtual environments and does not run to hide its nature and behavior.

In order to achieve good results in malware detection, and overcome signature-based detection and dynamic analysis weaknesses, many approaches to static analysis associated with machine learning have been investigated in recent works. Static analysis aims to study a file without running it to understand its purpose and nature. The most natural way is to extract features based on binary file bit statistics (entropy, distributions…) then to use ML algorithms to perform a binary classification (Random Forest, XGBoost, LightGBM for example). Among other things, the quality of detection models depends on features used for training and on the amount of data. In this regard, Anderson et al. \cite{anderson2018ember} provide Ember, a very good dataset to train ML algorithms. On the other hand, Raff et al. \cite{raff2017malware} use Natural Language Processing tools to analyse bit sequences extracted from binary files. Their MalConv algorithm gives very good results but requires a lot of computing power to train it. Moreover, it has recently been shown that this technique is very vulnerable to padding and GAN-based evasion methods. To overcome these weaknesses, Fleshman et al. \cite{fleshman2018non} developed Non-Negative MalConv which reduces the evasion rate but provides a slight drop in accuracy.

Natataj et al. \cite{nataraj2011malware} introduced the use of grayscale images to classify 25 malware families. Authors convert binary files into images and use the GIST algorithm to extract important features from them. They train a K-NN with these features and obtain a percentage of accuracy of $97.25$\%. In addition to presenting a good classification rate, this method has the characteristic of offering better resilience against obfuscation, especially against packing, the most used obfuscation method. In the continuity of this work, Vu et al. \cite{vu2019convolutional} proposed the use of RGB (for Red Green Blue) images for malware classification with their own transformation method called Hybrid Image Transformation (HIT). They encode the syntactic information in the green channel of an RGB image, while the red and blue channels capture the entropy information.

In view of the interest in image recognition, with ImageNet \cite{deng2009imagenet} for example, and performance improvements \cite{alom2018history} on the topic for several years, some authors proposed using a Convolutional Neural Network (CNN) apply to binary files converted into grayscale images for malware classification. Rezende \cite{rezende2017malicious} applied transfer learning on ResNet-50 for malware family classification and achieved a percentage of accuracy of $98.62$\%. To go deeper into the subject, Yakura et al. \cite{yakura2018malware} used attention mechanism with CNN to highlight areas in grayscale images that help with classification. Also, they relate areas of importance to the disassembled function of the code.

Another principal trend in malware research is to protect detection models against obfuscation techniques. Many malware are known, but they have been modified to make them undetectable. For example, polymorphic \cite{Sharma2014} and metamorphic \cite{Zhang2007} malicious files embed mechanisms that modify their code apparently but not their behavior. Moreover, malware authors can alter them manually. Kreuk et al. \cite{kreuk2018adversarial} inject bytes directly into the binary file to perturb the detection model without transforming its functions. Another modification is to pack malware, and it is one of the commonly used methods to easily evade antivirus software. Aghakhani et al. \cite{aghakhani2020malware} give an overview of the limits of detection models to spot packed malware.

\subsection{Contributions and paper plan}
The contributions of the study can be summarized as follows: 

\begin{itemize}
    
    \item Different detection methods are tested on a real database containing complex malware harvested in the company. In particular, we propose detection models that use grayscale image and HIT preprocessing on our own dataset of binary files. We compare the results of our models with models (LGBM, XGBoost, DNN) trained with the Ember dataset and preprocessing. 
    
    \item We propose models that take into consideration the fact that binary files can be packed or encrypted. One objective of this method is to reduce the false positive rate due to the interpretation of some models that modified files are necessarily malicious. Another objective is to give malware analysts a tool that provides them with more information on the nature of a suspicious file. 
    
    \item We implement attention mechanisms to interpret the results of our image recognition algorithms. This method is used to extract the parts of the image, and therefore of the binary, which contributed the most to the scoring of the classification. This allows this information to be passed on to security analysts to facilitate and accelerate reverse engineering work on the malware. Their feedback is then used to understand algorithm errors and improve this aspect. 
\end{itemize}

This paper is organized as follows: in Section \ref{Sec:DataPreproc}, we give a description of our dataset, discussing its advantages and the different preprocessing methods involved. In Section \ref{Sec:AlgoRes}, we present the different models trained on Ember or on our own datasets. We compare the models and discuss the results and performances. Section \ref{Sec:PackedAttention} is dedicated to the analysis of modified samples and to attention mechanisms, two methods which can be an interesting aid for analysts. Section \ref{Sec:Conclusion} summarizes the results and concludes the paper.

\section{Dataset and preprocessing}
\label{Sec:DataPreproc}
\subsection{Description of binaries dataset}

\begin{figure}[!ht]
    \centering
    \begin{subfigure}[h]{0.4\textwidth}
        \includegraphics[width=\textwidth]{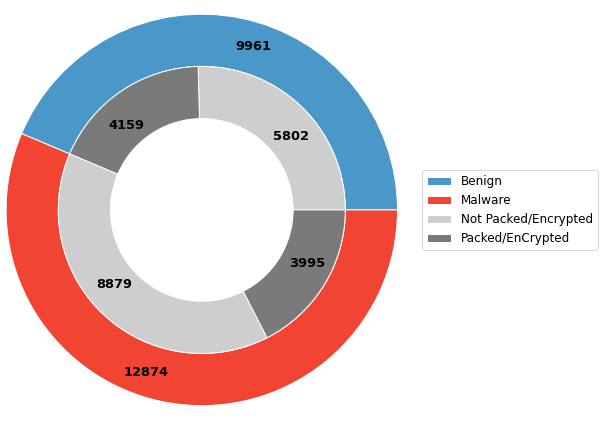}
    \end{subfigure}
    \caption{Distribution of our dataset}
    \label{fig:repartition}
\end{figure}

Our dataset contains 22,835 benign and malware in Portable Executable (PE) format, including packed or encrypted binary files. Fig. \ref{fig:repartition} shows the exact distribution of the dataset. The benign files are derived from harvested Windows executables, and the malwares have been collected in companies and on sandboxes. The dataset's main distinguishing feature is that these malware are relatively difficult to detect. As evidence, they have been undetected by some sandboxes or antivirus programs. As our dataset contains complex and non-generic malware, it should prevent overfitting during the training of our models.

To train machine learning algorithms, we use the Ember dataset which contains 600,000 PE files for training and we test on our own dataset to see the results. For the image-based algorithm, we split the dataset into $80 \%$ of training data, $10 \%$ of testing data and $10 \%$ of validation data. This distribution is the most optimized to keep a training sample large enough and a testing sample complex enough. 

\subsection{Is the malware modified ?}
\label{sec:bytehist}

A recurring problem when doing static analysis is the analysis of packed or encrypted executables, that we include under the term "modified" file in the rest of the paper. Artificial intelligence algorithms will often classify them as malicious even though many benign executables are modified for industrial or intellectual property reasons, for example. This is understandable given that these processes will drastically alter the entropy and distribution of bytes in the executable. A line of thought for better performance is to take into consideration the modifying nature of binary files during the training of detection models.  

\begin{figure}[!ht]
    \centering
    \begin{subfigure}[h]{0.20\textwidth}
        \includegraphics[width=\textwidth]{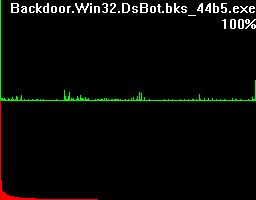}
        \caption{Malware not packed}
    \end{subfigure}
    \begin{subfigure}[h]{0.20\textwidth}
        \includegraphics[width=\textwidth]{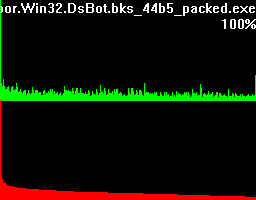}
        \caption{Malware packed}
    \end{subfigure}
    \begin{subfigure}[h]{0.20\textwidth}
        \includegraphics[width=\textwidth]{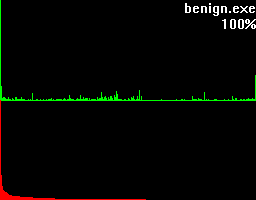}
        \caption{Benign not packed}
    \end{subfigure}
    \begin{subfigure}[h]{0.20\textwidth}
        \includegraphics[width=\textwidth]{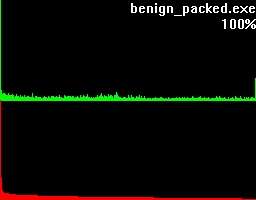}
        \caption{Benign packed}
    \end{subfigure}
    \caption{Byte distribution comparison between malware and benign with ByteHist}
    \label{fig:bytehist}
\end{figure}

Upstream of the analysis, the use of software such as ByteHist \cite{bytehist} gives an idea of the nature of a file. Indeed, ByteHist is a tool for generating byte-usage-histograms for all types of files with a special focus on binary executables in PE-format. ByteHist allows us to see the distribution of bytes in an executable. The more the executable is packed, the more uniform the distribution is. Fig. \ref{fig:bytehist} presents some byte distribution examples of one malware and one benign not packed and their UPX-transformed equivalents.

As we can see, UPX changes the byte distribution of binary files, in particular for the malware examples with more modifications than the benign file. Also, it is a common packer and it is easy to unpack binary files created with UPX. However, many malware are packed with more complex software, making analysis more difficult.

\subsection{Image-based malware transformation}

Before discussing how to turn a binary into an image, let us briefly explain why we use images. First of all, different sections of a binary can be easily seen when it is transformed into an image, so that it can give a first orientation to an analyst as to where to look, as we will see in the next section. Then, as we discussed in the introduction, malware authors can modify parts of their files or use polymorphism to change their signatures or produce recent variants. Images can capture small alterations yet retain the global structure of the malware.

Given a static binary, we map it directly to an array of integers between 0 and 255. Hence, each binary is converted into a one-dimensional array $v \in [0, 255]$, $v$ is then reshaped into a two-dimensional array and we follow the resizing scheme as presented in \cite{nataraj2011malware}. That is, the width is determined with respect to the file size. The height of the file is the total length of the one-dimensional array divided by the width. We rounded up the height and pad zeros if the width is undivisible by the file size. This method allows us to transform a binary into a grayscale image. The main advantage of this process is that it is very fast. For 20,000 binaries, it takes at most a few minutes.

\begin{figure}[!ht]
    \centering
    \begin{subfigure}[h]{0.2\textwidth}
        \includegraphics[width=\textwidth]{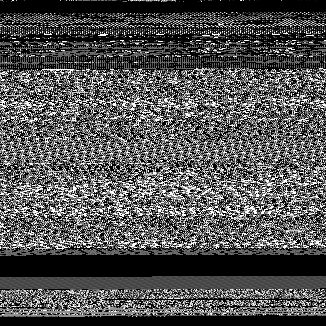}
        \caption{Malware not packed}
    \end{subfigure}
    \begin{subfigure}[h]{0.2\textwidth}
        \includegraphics[width=\textwidth]{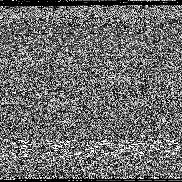}
        \caption{Malware packed}
    \end{subfigure}
    \begin{subfigure}[h]{0.20\textwidth}
        \includegraphics[width=\textwidth]{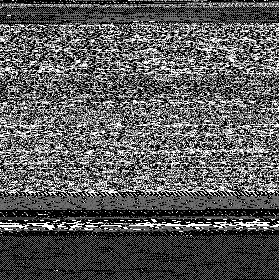}
        \caption{Benign not packed}
    \end{subfigure}
    \begin{subfigure}[h]{0.2\textwidth}
        \includegraphics[width=\textwidth]{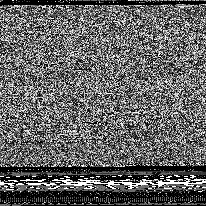}
        \caption{Benign packed}
    \end{subfigure}
    \label{lab:grayscale}
    \caption{Grayscale representation of some binary files}
\end{figure}

Vu et al. \cite{vu2019convolutional} give different methods to transform a binary into an RGB image.
Their color encoding scheme is based on the fact that green is the most sensitive to human vision and has the highest coefficient value in image grayscale conversion. In particular, with their HIT method,  they encode the syntactic information into the green channel of an RGB image, while the red and blue channels capture the entropy information. In this way, clean files will intuitively have more green pixels than malicious files, which contain higher entropy with higher red/blue values. This transformation gives very good results with image recognition algorithms. The only downside is the transformation time. It takes an average of 25 seconds to transform a binary into an image with their HIT method. 

\begin{figure}
    \centering
    \begin{subfigure}[h]{0.2\textwidth}
        \includegraphics[width=\textwidth]{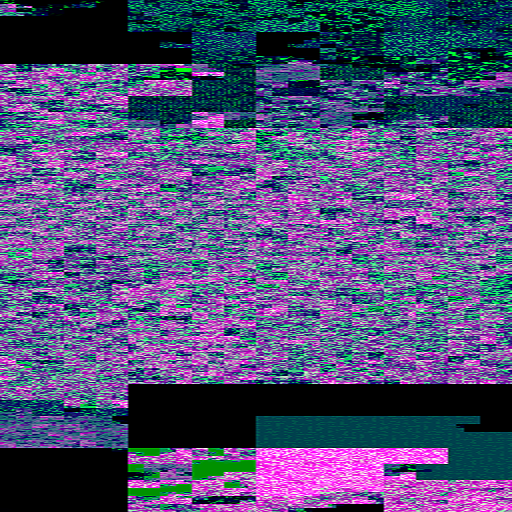}
        \caption{Malware not packed}
    \end{subfigure}
    \begin{subfigure}[h]{0.2\textwidth}
        \includegraphics[width=\textwidth]{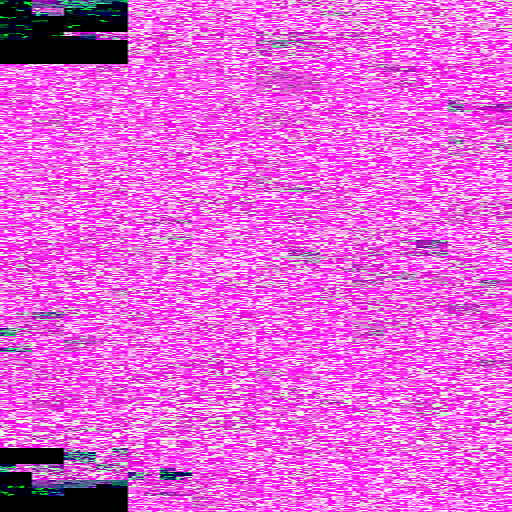}
        \caption{Malware packed}
    \end{subfigure}
    \begin{subfigure}[h]{0.2\textwidth}
        \includegraphics[width=\textwidth]{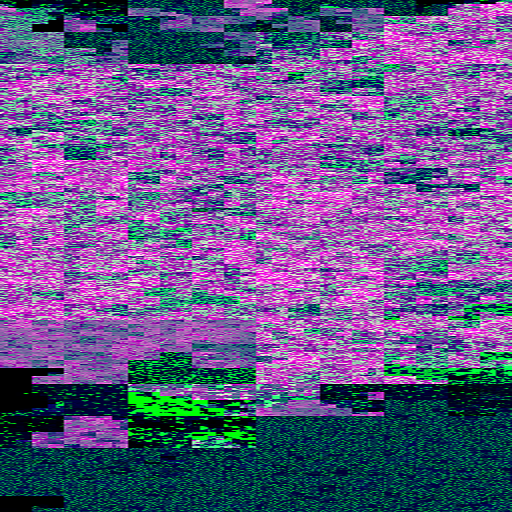}
        \caption{Benign not packed}
    \end{subfigure}
    \begin{subfigure}[h]{0.2\textwidth}
        \includegraphics[width=\textwidth]{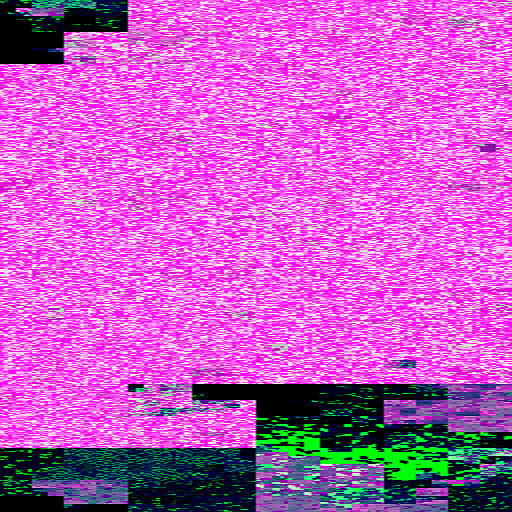}
        \caption{Benign packed}
    \end{subfigure}
    \caption{HIT representation of some binary files}
    \label{fig:grayscales-hit}
\end{figure}

Fig. \ref{fig:grayscales-hit} presents grayscale and HIT transformations of the binary file introduced previously. 

\section{Detection based on statics methods}
\label{Sec:AlgoRes}

In this part, we study and compare three approaches to malware detection based on static methods and machine learning algorithms:

\begin{itemize}
    \item First, we train three models on the Ember dataset with their own feature extraction method. 
    \item Then, using this time grayscale images as input, we propose a CNN to detect malware and, to go further, three hybrid models. 
    \item Finally, we train another CNN on an RGB image using the HIT method.
\end{itemize}

\subsection{Algorithms on binary files}
\label{staticanlysis}

\begin{figure}[!ht]
    \centering
    \begin{subfigure}[h]{0.5\textwidth}
        \includegraphics[width=\textwidth]{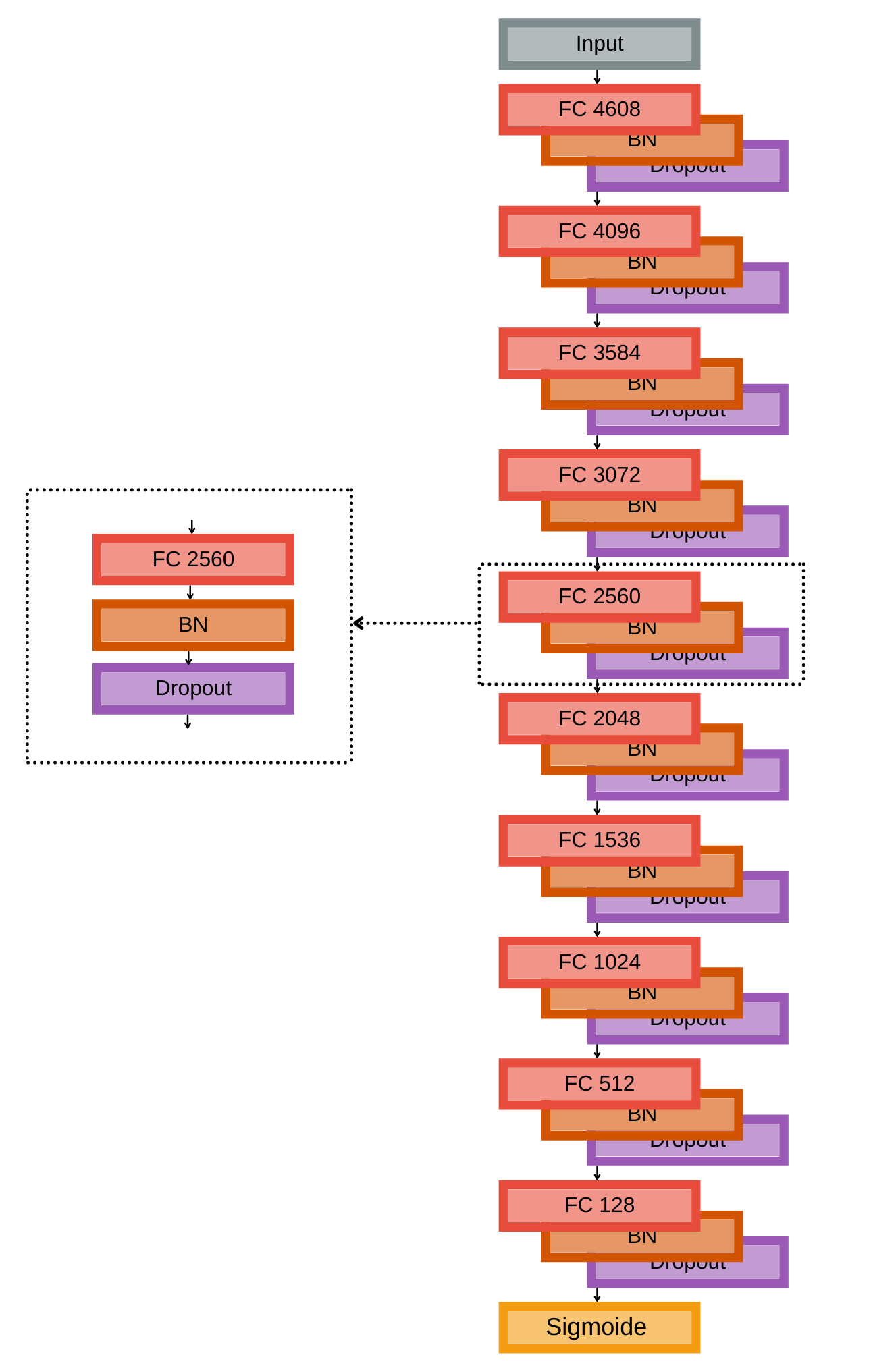}
    \end{subfigure}
    \caption{DNN architecture}
    \label{fig:DNN_archi}
\end{figure}

For the static analysis, we will test three algorithms: XGBoost, LightGBM and a deep neural network (DNN) whose architecture is presented in Fig. \ref{fig:DNN_archi}. XGBoost \cite{chen2016xgboost} is a reference algorithm for testing data but, on a large dataset, there can be some issues with computing time. That's why we also compare it with LightBGM \cite{ke2017lightgbm} which is used by Ember in connection with their dataset. 

Let us quickly introduce the LightGBM algorithm which is still less known. It uses a novel technique of Gradient-based One-Side Sampling (GOSS) to filter out the data instances to find a split value while XGBoost uses a pre-sorted algorithm and a histogram-based algorithm for computing the best split. Here, instances are observations. Its main advantages compared to other algorithms like Random Forest or XGBoost are:

\begin{itemize}
    \item Faster training speed and higher efficiency.
    \item Lower memory usage (replaces continuous values with discrete bins, which results in lower memory usage).
    \item Better accuracy with more complex tree.
\end{itemize}

Specifically, if we focus on it in this study, it is mainly because of its capacity to handle a huge amount of data. It can perform equally well with large datasets and present a significant reduction in training time as compared to XGBoost.

To begin with, we train algorithms XGBoost and LightGBM on the Ember dataset, and we test them on our own data. In addition, we train a DNN on the Ember learning dataset too, because this kind of models goes hand to hand with a large dataset that contains so many features. We use the F1 score and accuracy score to compare models between them. Results are collected in Table \ref{table:static}.

\begin{table}[H]
    \centering
    \caption{Static models F1 score and accuracy}
    \label{table:static}
    \begin{tabular}{|c|c|c|}
    \hline
        & F1 Score & Accuracy Score  \\
    \hline
        LGBM & 0.9110 & 0.9001 \\
    \hline
        XGBoost & 0.8275 & 0.7748 \\
    \hline 
        DNN & \textbf{0.9160} & \textbf{0.9071} \\
    \hline
    \end{tabular}
\end{table}

From this table, we can see that LightGBM and DNN performances are very close but that XGBoost is less performant (either in precision or computing time). 

\subsection{Algorithms on grayscale images} 
\label{Sec:grasycale}

\begin{figure}[!ht]
    \centering
    \begin{subfigure}[h]{0.5\textwidth}
        \includegraphics[width=\textwidth]{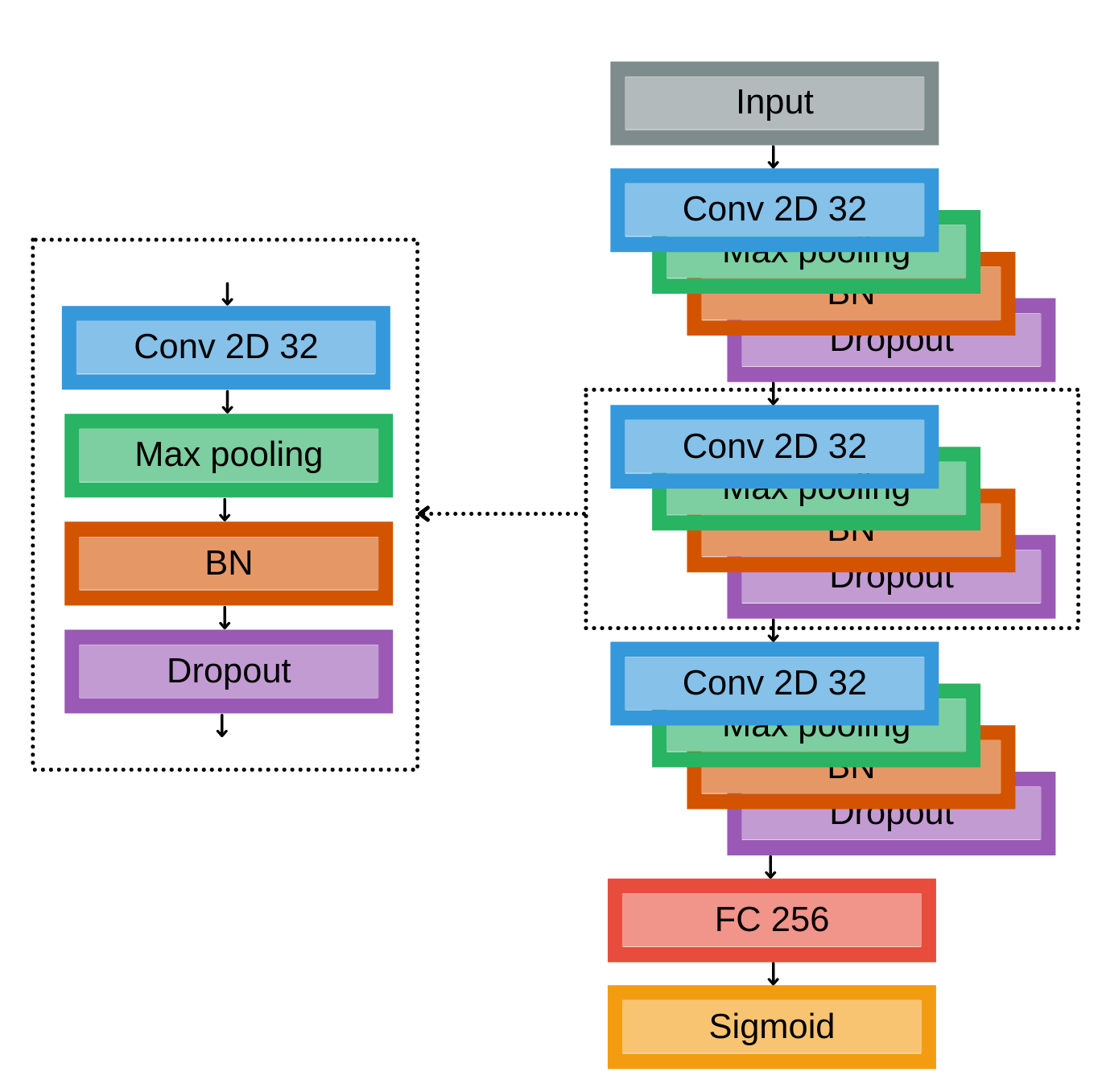}
    \end{subfigure}
    \caption{CNN architecture}
    \label{CNN_archi}
\end{figure}

Based on Nataraj et al. \cite{nataraj2011malware} work, we transform our dataset into grayscale images and employ them to train CNN. Our CNN is composed of three convolutional layers, a dense layer with a ReLU activation function and a sigmoid function for scoring binaries as presented in the Fig. \ref{CNN_archi}. Also, inspired by \cite{xiao2019intrusion}, we propose hybrid models combining CNN and LightGBM, RF or Support Vector Machine (SVM). Firstly, we use CNN to reduce the number of dimensions and, for each binary image, we go from 4,096 features to 256. Then, we use these 256 new features to train RF, LightGBM and SVM models.
As shown in Table \ref{table:grayscale}, F1 and accuracy scores are still used to compare models.

\begin{table}[H]
        \centering
        \caption{Grayscale models F1 score and accuracy}
        \label{table:grayscale}
        \begin{tabular}{ |c|c|c|}
        \hline
         & F1 Score & Accuracy Score  \\
        \hline
        CNN& 0.8786 & 0.8703\\
        \hline
        CNN+LGBM&  0.8827&  0.8703\\
        \hline
        CNN+RF&  \textbf{0.8914}&  \textbf{0.8804}\\
        \hline
        CNN+SVM&  0.8895&  0.8791\\
        \hline
        \end{tabular}
\end{table}

As can be seen, the hybrid model combining CNN and RF outperforms the four grayscale models, but the overall results are close. Also, the performances are relatively close to those of the LightGBM and the DNN presented in section \ref{staticanlysis}.It should be noted that the grayscale models are trained using only 19,400 binary files, whereas the previous models' training set consists of 600,000 binary files. So, with the grayscale transformation and a dataset thirty times smaller, our grayscale models remain reliable for malware detection compared to conventional models and preprocessing.

\subsection{Algorithms on RGB images}

We are now evaluating our CNN on the basis of RGB images using HIT transformation. Table \ref{table:HIT} shows F1 and accuracy scores on the test sample. 

Even if the performance of the RGB model is better than the others previously presented, training on a local machine is quite long with RGB images, but scoring on a single one is fast. Due to the complexity of HIT algorithm, the transformation time of binaries into images is quite long and takes, on average, 25 seconds for a sample, against less than one second for the grayscale transformation. First of all, it increases the learning process considerably if we include the time to convert the 24,000 samples. Moreover, when predicting malware, the score is obtained in less than one second, but the time for converting the binary into an image is added. Considering this, it’s useless to use HIT transformation compared to grayscale transformation in corporate case, this is why we will not dwell on training other models with  HIT preprocessing. 

\begin{table}[H]
    \centering
        \caption{RGB CNN model F1 score and accuracy}
        \label{table:HIT}
        \begin{tabular}{ |c|c|c|}
            \hline
             & F1 Score & Accuracy Score  \\
            \hline
            CNN& 0.934 & 0.94\\
            \hline
        \end{tabular}
\end{table}

\section{Modified binaries analysis and attention mechanism}
\label{Sec:PackedAttention}

The objective, in addition to having the most accurate results possible, is to make them usable by an analyst. To do this, we must be able to understand why our algorithms give high scores or not. This allows us to improve the learning process in case of error, but also to give a pointer to the analysts to know where to look. We propose two approaches to facilitate the understanding and analysis of malware: 

\begin{itemize}
    \item The first approach is to use, during the training of our algorithms, information about the nature of binary files. In particular, we know if the binary files of the training set are modified. The purpose is to reduce false positive results caused by these two obfuscation methods, and also give more information about the nature of the new suspect files. 
    \item The second approach is the use of an attention mechanism on model trains with grayscale images. We can generate heatmap using the attention mechanism to detect suspicious patterns. Furthermore, the heatmap also helps to understand the results of the malware detection model. 
\end{itemize}

\subsection{Modified binaries}

In order to reduce the false positive rate due to obfuscation, we also provide two models which are trained while taking into account the altered nature of the binary file. The two models take in input grayscale images. 

\begin{enumerate}
    \item The first model is a CNN which returns  output information on the nature of the binary file, if it is a malware or not, and if it is obfuscated or not. So, with a single CNN, we have double knowledge on the characteristics of the binary file. This model achieves a F1 score of 0.8924 and an accuracy score of 0.8852.
    \item The second model is a superposition of three CNNs. The first one is used to separate binary files, according to whether they are obfuscated or not, with an accuracy of $85\%$. The two others are used to predict if a binary file is a malware or a benign and each model is respectively trained on modified binary file and not modified binary file. The main advantage of this model is that each CNN is independent from the other two and can be retrained separately. They also use different architectures to improve the generalization of the data used to train them. We get a F1 score of 0.8797 and an accuracy score of 0.8699 for this model.
\end{enumerate}

As we can see, the first model gives better results than the second model. Also, it can determine if a binary file is modified with an accuracy rate of $84\%$. This information could  help malware analysts to have a better expertise. For example, this can explain why some benign files are detected as malware. Moreover, it can encourage the use of sandboxes for certain suspicious files if they modified and if the result of the malware detection is ambiguous.

\subsection{Interpretibility of results and most important bytes}

\begin{figure}[!ht]
    \centering
    \begin{subfigure}[h]{0.6\textwidth}
        \includegraphics[width=\textwidth]{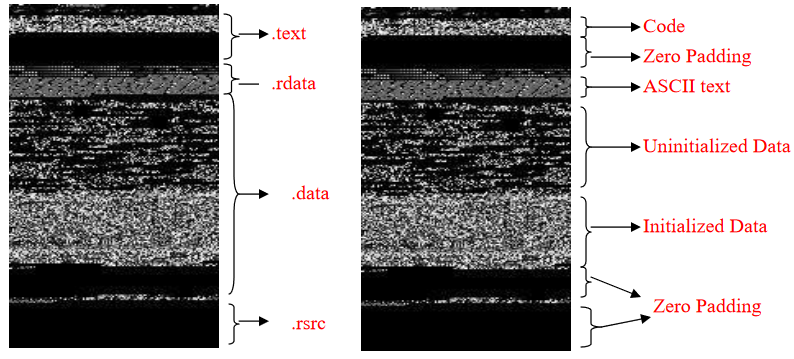}
    \end{subfigure}
    \caption{Grayscale image with corresponding section (left) and texture interpretation (right)}
    \label{fig:PEgrayscale}
\end{figure}

In this section, we present an approach that can help analysts to interpret the results of detection models based on the transformation of binary files into grayscale images.
A grayscale image representation of an executable has distinct texture depending on the relevant sections of the file \cite{conti2010visual}. We can use tools like PE file to extract information from the binary and see the relationship between the PE file and its grayscale image representation. Fig. \ref{fig:PEgrayscale} shows an executable transformed into an image, the corresponding sections of the PE file (left), and information associated with each texture (right).

The associations of PE file and grayscale image can allow an analyst to quickly visualize the areas of interest of the binary file. To go further in the analysis, it should be necessary to study which parts of the image have contributed to the results of malware detection algorithm. To do this, we use attention mechanisms that consist of highlighting the pixels that influence the most the prediction score of our algorithm.

\begin{figure}[!ht]
    \centering
    \begin{subfigure}[h]{0.2\textwidth}
        \includegraphics[width=\textwidth]{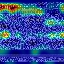}
        \caption{Malware not packed}
    \end{subfigure}
    \begin{subfigure}[h]{0.2\textwidth}
        \includegraphics[width=\textwidth]{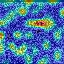}
        \caption{Malware packed}
    \end{subfigure}
    \begin{subfigure}[h]{0.2\textwidth}
        \includegraphics[width=\textwidth]{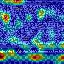}
        \caption{Benign not packed}
    \end{subfigure}
    \begin{subfigure}[h]{0.2\textwidth}
        \includegraphics[width=\textwidth]{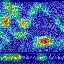}
        \caption{Benign packed}
    \end{subfigure}
    \caption{Attention map of some binary files}
    \label{fig:gradcam}
\end{figure}

We use GradCAM++ \cite{chattopadhay2018grad} with our own CNN presented in section \ref{Sec:grasycale}. The GradeCAM++ algorithm extracts from the CNN the pixels that influence the most the model's decision, i.e. those that determine if the file is benign or malicious. It returns heatmap which could be interpreted as follows, the more the image area impacts the CNN prediction, the warmer the coloring. Fig. \ref{fig:gradcam} presents heatmaps of the four binaries introduced in section \ref{sec:bytehist}. We observe that the malware and its packed version do not have the same activation zones. We can make the same remark about the benign. Also, as the byte distribution of the packed malware has undergone a lot of modifications, we see that more zones and pixels are lit up compared to the original malware. This means that the model needs more information to make a decision.

\begin{figure}[!ht]
    \centering
    \begin{subfigure}[h]{0.2\textwidth}
        \includegraphics[width=\textwidth]{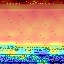}
        \caption{Benign}
    \end{subfigure}
    \begin{subfigure}[h]{0.2\textwidth}
        \includegraphics[width=\textwidth]{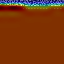}
        \caption{Malware}
    \end{subfigure}
    \caption{Example of binary files detected as malware due to padding}
    \label{fig:padding}
\end{figure}

On the other hand, this kind of representation could help to understand why a binary file is misclassified. For example, a common evasion technique is called padding and relies on adding byte sequences to artificially increase the size of a file and fool the antivirus. With image representation, this kind of technique is easily detected. However, as we can see in Fig. \ref{fig:padding}, for the two examples, the padding zone is considered as an important part of the file. Even if the malware is correctly detected, the benign file is misclassified and labeled as malware. So, the padding is considered as a malicious criterion. This knowledge could be taken into consideration to increase the performance of the malware detection model.

The activation map on binary file images seems to be an interesting and useful tool to help malware analysts. However, it is necessary to deepen this subject to fully exploit the potential of this technique. Indeed, we show the possible use of heatmap for packing or padding binary files analysis but there are many other obfuscation techniques. Furthermore, we focus here on images of malware and benign, but an extension of this approach will be to extract code directly from the binary file based on the hot zone of the corresponding heatmap.

\section{Conclusion and results}
\label{Sec:Conclusion}

Before concluding this paper, let us start by summarizing the results. In fact, CNN trained on RGB images based on HIT provides better results. However, the transformation time is too important to provide an efficient method for industrial deployment. Next, the DNN and LightGBM models demonstrate the expected effectiveness of the Ember dataset. Our models, which use grayscale images as input, are slightly less efficient than theirs. The results, however, are quite comparable to a training sample thirty times smaller. Finally, the two models train on grayscale images with information on whether the original binary file has undergone modifications or not pointed out the potential of this method for malware detection. They also provide more information on binary files than common detection models. CNN algorithm hybridized with RF, LGBM and SVM show an interesting detection potential. We will focus on this in future work to determine capacity or limit of this kind of models. 

We have presented in this article different approaches to static malware detection. A recurring problem in many organizations is the computational time required for dynamic analysis of malware in a sandbox. However, with some modified malware, we know it is sometimes the only solution. We do not claim here to be able to replace this analysis, but we prefer proposing an overlay. Our algorithms enable us to quickly analyze a large amount of malware, determine which ones may or may not be malicious, while moderating the result if the binary is detected as modified or not. This can allow us to use dynamic analysis only on those binaries and, thus, save time and resources. In addition, analyzing the most important pixels and sections in image can also provide significant information for analysts. That can save them time in the detailed study of the suspicious binary by indicating where to look for it.

In future work, we will concentrate on attention mechanisms. The objective is to match the areas of importance, extracted with attention mechanisms, with the associated malicious code to help the analysts in their work. On the other hand, we want to use reinforcement learning to understand and prevent malware evasion mechanisms.

\vspace{12pt}

\end{document}